%% file: paper.tex
\documentclass[a4paper,nofootinbib,pra,twocolumn,twoside,superscriptaddress]{revtex4-1}

\include{macros}

\usepackage{graphicx}
\usepackage{hyperref}
\hypersetup{ 
  colorlinks=true,
  linkcolor=blue,
  citecolor=blue,
  urlcolor=blue,
  breaklinks=true}
\usepackage{ifpdf}
\ifpdf
\fi

\newcommand\CHANGED[1]{#1}

\begin{document}

\title{On-Line Estimation of \CHANGED{Local Oscillator} Noise and Optimisation of Servo Parameters in Atomic Clocks}

\newcommand*\QUEST{
QUEST Institute for Experimental Quantum Metrology, Physikalisch-Technische Bundesanstalt, 38116 Braunschweig, Germany}
\author{Ian D. Leroux} \email{Ian.Leroux@nrc-cnrc.gc.ca}
  \altaffiliation[Current Address: ]{\CHANGED{National Research Council Canada, Ottawa, Ontario, Canada K1A 0R6}}
  \affiliation{\QUEST}
\author{Nils Scharnhorst} \affiliation{\QUEST}
\author{Stephan Hannig} \affiliation{\QUEST}
\author{Johannes Kramer} \affiliation{\QUEST}
\author{Lennart Pelzer} \affiliation{\QUEST}
\author{Mariia Stepanova} \affiliation{\QUEST}
\author{Piet O. Schmidt} \affiliation{\QUEST}
  \affiliation{Institut für Quantenoptik, Leibniz Universität Hannover, 30167 Hannover, Germany}
\date{\today}
\begin{abstract}
For atomic frequency standards in which fluctuations of the local oscillator \CHANGED{(LO)} frequency are the dominant noise source,
we examine the role of the servo algorithm that predicts and corrects these frequency fluctuations.
\CHANGED{We derive the optimal linear prediction algorithm,
showing how to measure the relevant spectral properties of the noise and optimise servo parameters
while the standard is running,
using only the atomic error signal.
We find that, for realistic LO noise spectra,
a conventional integrating servo with a properly chosen gain performs nearly as well as the optimal linear predictor.}
Using simple analytical models and numerical simulations,
we establish optimum probe times
as a function of clock atom number and of the dominant noise type in the local oscillator.
\CHANGED{We calculate the resulting LO-dependent scaling of achievable clock stability with atom number
for product states as well as for maximally-correlated states.}
\end{abstract}
\maketitle

The instability of frequency standards limits the total uncertainty achievable in a measurement of finite duration~\cite{ludlow_optical_2015, poli_optical_2013}.
This limit can be practically relevant even when performing measurements of static frequency ratios,
since many-month-long measurement campaigns place stringent demands on the reliability of all components in an experiment.
Instability becomes a fundamental concern when attempting to measure time-varying frequency ratios.
For instance, in the emerging field of chronometric leveling~\cite{vermeer_chronometric_1983, bjerhammar_relativistic_1985, delva_atomic_2013}, direct observation of tidal fluctuations expected in the gravitational red shift~\cite{lisdat_clock_2016} requires frequency ratio measurements with a fractional uncertainty at the level of \num{e-18} to be completed in a matter of hours.
Physics beyond the Standard Model might be detectable in clock frequency ratio measurements as postulated transient shifts associated with dark-matter domain walls~\cite{Derevianko2014} or ultralight scalar dark-matter candidates~\cite{van_tilburg_search_2015, stadnik_improved_2016}.
\CHANGED{Searches for such signals require the highest possible measurement resolution at timescales where the statistical uncertainty due to instability plays a far greater role than long-term systematic uncertainty.}

Of the noise processes contributing to the instability of atomic frequency standards, the most fundamental one is quantum projection noise~\cite{Itano1993}, which arises from the discreteness in the measurement results obtainable from a finite number of atoms. For an ensemble of $N$ uncorrelated two-level atoms, this noise imposes a minimum statistical uncertainty
\begin{equation}
  \qpnnoise = \frac{1}{\sqrt{N}}
  \label{eqn:defineqpn}
\end{equation}
on any measurement of the phase accumulated in an atomic superposition state.
\CHANGED{For a standard operating at a frequency $\wnom$
and in the ideal case of Ramsey interrogation without technical noise,
this leads to a long-term fractional instability~\cite{riehle_frequency_2004}}
\begin{equation}
  \clkadev\of\tau = \frac{1}{\wnom\Tprobe\sqrt{N}}\sqrt{\frac{\Tcyc}{\tau}}
  \label{eqn:definesql}
\end{equation}
where $\Tprobe$ is the duration of a single Ramsey interrogation and $\Tcyc$ is the length of the frequency standard's operating cycle, such that $\tau/\Tcyc$ measurements can be performed in an averaging time $\tau$. This quantum projection noise limit (QPN)%
\footnote{sometimes referred to as the Standard Quantum Limit (SQL)}
for clocks using uncorrelated atoms depends on the experimenter's choice of probe time $\Tprobe$,
becoming arbitrarily small for sufficiently long probe times.
Thus, Eqn.~\ref{eqn:definesql} sets no limit on achievable clock instability at long averaging times unless some additional scale in the problem restricts the choice of $\Tprobe$.

One such restriction is set by excited-state decay in the atoms,
which sets a fundamental limit to interrogation times.
The performance of optical frequency standards operating at this limit has been analysed in Refs.~\cite{Huelga1997, riis_optimum_2004, champenois_evaluation_2004, Peik2006}.
However, for many of the optical frequency standards now being investigated,
frequency fluctuations of the local oscillator restrict $\Tprobe$ to less than a second even when the atoms' excited-state lifetime is measured in minutes (\Sres) or even years (\Ybion)~\cite{ludlow_optical_2015}.
Because the local oscillator's noise is common to all the atoms in the standard,
and because it typically exhibits significant power-law temporal correlations,
its effects are qualitatively different from those of excited-state decay.
In fact, it might at first glance seem odd that local-oscillator noise limits clock stability at all:
the local oscillator frequency is in some sense the measurand in an atomic frequency standard
and its fluctuations are constantly monitored and corrected.
Local-oscillator noise affects the stability of the standard only to the extent that it cannot be corrected
by feedback from the atoms.
This can happen, for instance, if the cyclic atomic interrogation protocol allows undetected aliased frequency components of the local-oscillator noise spectrum to contaminate the output signal of the standard,
a phenomenon known as the Dick effect~\cite{Dick1987}.
Even in the absence of the Dick effect, however,
the quantised measurement signal from the atoms has a fundamentally limited dynamic range:
one cannot extract more than $\log_2\of{N+1}$ bits of frequency information
from a single measurement of $N$ atoms~\cite{Buzek1999}.
The useful domain of the measurement,
i.e. the frequency band in which it can be unambiguously interpreted,
must be broad enough to cover the frequencies which the local oscillator is likely to emit in the interrogation.
Frequency excursions beyond the domain for which the reference provides useful information lead to less informative measurement results, and hence to degraded instability.
In the worst case the servo,
working from ambiguous or uninformative measurement results,
may be unable to keep the output frequency locked to the atomic reference.
The output frequency then either hops between different zero-crossings of a frequency-periodic Ramsey error signal
or drifts aimlessly far from the resonance of a Rabi error signal.
This case is catastrophic and the operating parameters must be chosen to make it vanishingly unlikely.
Thus, even in the absence of the Dick effect (e.g. with dead-time-free Ramsey interrogation~\cite{biedermann_zero-dead-time_2013}),
the achievable measurement resolution ultimately depends on the scale of local-oscillator frequency fluctuations seen by the atoms, and hence on the performance of the clock's feedback loop which corrects these fluctuations.

In this work, we study the limits to the stability of frequency standards dominated by local-oscillator noise with realistic temporal correlations.
We focus on clocks using a single ensemble of atoms periodically interrogated using the same protocol for every interrogation cycle,
whose instability we quantify using the Allan variance at long times.
Our work is thus less general, but more directly relevant to current experiments,
than analyses of multi-ensemble clocks or of interrogation protocols which are modified on-the-fly~\cite{Rosenband2013,Kessler2014,Mullan2014,Chabuda2016},
and our approach is a more concrete complement to the derivation of universal performance bounds in mathematically idealised settings~\cite{Chabuda2016,fraas_analysis_2016}.
Using simple analytical arguments and numerical simulations of clocks with different local-oscillator noise spectra,
we study the performance of the servo controller which predicts and corrects local-oscillator noise and then analyse its implications.
After establishing notation and conventions in Sec.~\ref{sec:notation},
we begin by deriving the optimal linear prediction algorithm and evaluating its performance in Sec.~\ref{sec:ctrldesign}.
In Sec.~\ref{sec:online} we show that
a feedback controller with near-optimal performance can be designed without prior knowledge of the noise spectrum,
by monitoring the error signals in normal clock operation.
We also show that the same techniques provide useful diagnostic information on the local oscillator's noise,
allowing on-line monitoring of its performance.
We turn to the effects of the noise in Sec.~\ref{sec:ltstab},
in which we derive a modification to the QPN that takes into account the performance of the servo controller.
This modified QPN formula predicts an overall limit to achievable clock instability,
which is attained for an optimal choice of atomic interrogation time that we discuss in Sec.~\ref{sec:refparams}.
Section~\ref{sec:entanglement} considers the merits of using entangled atomic states to modify the phase resolution of Eqn.~\ref{eqn:defineqpn},
giving a simple dimensional argument for the disappointing performance of maximally-correlated states in atomic clocks
and arguing for the superiority of states that enhance the dynamic range of atomic measurements~\cite{Kitagawa1993,Wineland1992,Wineland1994,Buzek1999}.
This result is complementary to that of Ref.~\cite{Huelga1997},
which considered independent dephasing of the atoms rather than the collective dephasing associated with the LO,
and takes into account temporal correlations in the LO noise rather than assuming a white spectrum as in Ref.~\cite{Borregard2013,Kessler2014}.
Section~\ref{sec:dblint} considers the instability of the clock at short times,
which may be limited by finite feedback gain rather than measurement noise,
showing that the second integrator recommended in Ref.~\cite{Peik2006}
to correct for linear drift of the LO
is also necessary to saturate the QPN limit in the presence of random-walk noise.
We conclude with some remarks on proposed frequency standards whose design does not follow the conventional pattern we consider in this work.

\section{Setup and notation}
\label{sec:notation}

\begin{figure}
  \centering
  \includegraphics{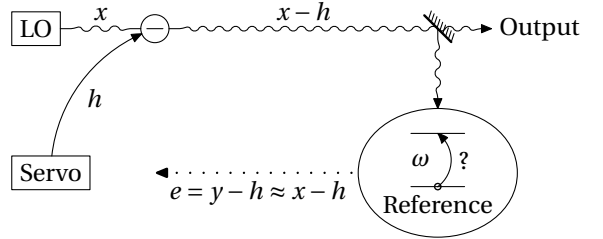}
  \caption{
    General model of a periodically-stabilised atomic clock.
    A local oscillator (LO) emits a signal with a fractional deviation $\lofreq$ from the nominal clock frequency.
    The servo controller attempts to predict $\lofreq$ and corrects the frequency by its prediction $\predfreq$.
    The corrected signal (with fractional frequency deviation $\lofreq - \predfreq$) is then used to interrogate an atomic reference,
    which produces an estimate $\errsig$ of the prediction error or, equivalently,
    an estimate $\estfreq$ of the LO's unknown frequency deviation $\lofreq$.
    These estimates can be used by the servo in future predictions.
    } \label{fig:setup}
\end{figure}

Figure~\ref{fig:setup} sketches the structure of the frequency standards we consider
and summarises the notation we will use for the various signals we must consider.
The goal of the standard is to produce a continuous classical oscillatory signal whose frequency corresponds to a reference transition frequency $\wnom$ in a particular atomic species.
As the classical signal is generated by a macroscopic local oscillator (LO) subject to environmental perturbations,
its frequency $\wlo$ will differ from the target frequency $\wnom$ by a fluctuating fractional discrepancy $\lofreq = (\wlo / \wnom) - 1$.
The scale of these fluctuations is summarised in the Allan deviation $\loadev\of\tau$ of the LO.
In order to suppress these fluctuations,
a servo controller generates a prediction $\predfreq$ of the LO frequency error,
which is used to frequency-shift the LO output signal back to atomic resonance.
The resulting signal, with net fractional frequency error $\lofreq-\predfreq$,
is provided to the users of the standard and has an Allan deviation $\clkadev\of\tau$.
This corrected signal is supplied to a reference,
where it interacts with $N$ atoms according to some fixed interrogation protocol,
such as Ramsey or Rabi interrogation.
Measurement of the atoms' state at the end of the interrogation protocol conveys some information on the residual frequency error $\lofreq - \predfreq$,
which we express as an error estimate $\errsig$.
The error estimate might, for instance, correspond to the imbalance of atomic state populations at the end of a Ramsey sequence divided by the accumulated phase $\omega \Tprobe$.
For consistency, we express the error estimate in the same units as $\lofreq$ and $\predfreq$,
so that $\estfreq = \predfreq + \errsig$ is an estimate of the LO's uncorrected frequency error $\lofreq$,
one that uses only the most recent atomic data and takes no account of previous measurements.
Note that $\estfreq$ and $\errsig$ differ from,
and fluctuate more than,
 $\lofreq$ and $\lofreq - \predfreq$ respectively,
because they are affected by the noise of the atomic reference.

We consider periodically-stabilised frequency standards with an operating cycle of period $\Tcyc$,
where the reference provides a series of error estimates $\{\errsig_i\}$.
At any given point in time, we label them as follows:
$\errsig_1$ is the most recent available error estimate,
$\errsig_2$ the preceding error estimate, and so forth.
$\errsig_0$ is then the error estimate which will be produced at the end of the current operating cycle.
We label the other signals similarly:
$\predfreq_0$ is the servo's prediction of the (average) LO frequency error $\lofreq_0$ during the current operating cycle,
while $\predfreq_j$ and $\lofreq_j$ correspond to the $j$th most recent completed cycle.
Causality requires that the servo compute the prediction $\predfreq_0$ without knowledge of $\errsig_0$,
using only $\{\errsig_1, \errsig_2, \ldots\}$ or, equivalently, $\{\estfreq_1, \estfreq_2, \ldots\}$.

The Allan deviation $\clkadev\of\tau$ is that of the physical signal produced by the frequency standard as it is operating.
With the exception of Sec.~\ref{sec:dblint},
most of the analysis presented in this work also applies to ``paper clocks'',
i.e. virtual signals generated by post-processing measurement data.
Although the post-processing need not respect causality and can use later measurements to correct estimates of the frequency at earlier times,
the quality of the measurements themselves still depends on the ability of the (causality-respecting) servo to keep the corrected LO frequency near the atomic resonance frequency $\wnom$ while the clock is running,
and constraints on this ability affect the performance of the reference no matter how the resulting data is subsequently used.

Where it is necessary to assume a definite interrogation protocol in the atomic reference,
we will focus on dead-time-free Ramsey interrogation,
where the measured signal depends on the average of the corrected signal frequency
during some interrogation time $\Tprobe$.
While we assume in our examples that $\Tprobe$,
which sets the frequency resolution of the interrogation,
is equal to $\Tcyc$, which sets the repetition rate of the interrogation cycle,
the two times are conceptually distinct and we will use separate symbols for them throughout.
References whose operating cycle includes dead time ($\Tprobe < \Tcyc$) or which use a different interrogation protocol
(such as Rabi or hyper-Ramsey~\cite{huntemann_generalized_2012, yudin_hyper-ramsey_2010})
will suffer from the Dick effect,
which can be modelled as additional measurement noise in the atomic reference.

In numerical examples we will consider LOs with simple power-law noise,
such that $\loadev^2\of\tau \propto \tau^\noisetype$,
with $\noisetype = -1, 0, 1$ for white frequency noise, flicker frequency noise and random walk of frequency noise, respectively.
As argued in the introduction, the LO noise gives the problem a characteristic time scale which ultimately limits the useful resolution of measurements on the atoms.
We define this time $\Tnought$, without assuming a particular form of LO noise spectrum,
by the implicit equation 
\begin{equation}
  \loadev\of\Tcycnought \wnom \Tnought = \SI{1}{\radian},
  \label{eqn:defineTnought}
\end{equation}
where $\Tcycnought$ is the cycle time of the clock when operated with a probe time $\Tnought$.
In other words, $\Tnought$ is the choice of probe time for which the LO Allan deviation at one clock cycle
is as large as the quantum projection noise of a single atom (Eqn.~\ref{eqn:definesql} with $N=1$).
This definition lets us combine the LO noise and the choice of probe time into a single dimensionless parameter
$\Tprobe / \Tnought$ which can be compared between clocks of different types using LOs with different performance.
Note that $\Tnought$ will be on the order of a few seconds for a typical current optical frequency standard with a fractional LO instability around \num{e-16}.

\CHANGED{In the remainder of this paper,
we will have frequent recourse to Monte-Carlo simulations of clocks.
Because our model assumes a fixed interrogation protocol,
it is possible to predetermine the start and end times of every radiation pulse in a simulated run of the clock,
and thus to generate efficiently the (noisy) mean frequency of the free-running LO during each pulse.
Given such a frequency history for the free-running LO,
it is straightforward to simulate the response of the atomic reference at each clock cycle
and the resulting servo correction for the next clock cycle.
White noise is generated as a random variable whose variance scales inversely with the duration of each pulse.
(Damped) random walks are obtained by first generating the frequency at the beginning and end of each pulse as a (damped) running sum of steps whose variance depends on the time step length,
then computing the expectation value of the mean of the random walk in each pulse given fixed start- and end-points,
and finally adding a white noise component corresponding to the dispersion of the mean about this expectation value.
Flicker-frequency noise is generated as a sum of damped random walks with damping time constants ranging by factors of 2 from \SI{1}{\percent} of the shortest pulse in the clock's operating cycle (the shortest time scale in the problem) up to 100 times the duration of the entire run (the longest time scale in the problem).}

\section{Servo Controller Design}
\label{sec:ctrldesign}

We now focus our attention on the servo.
Given a history $\ldots, \predfreq_3, \predfreq_2, \predfreq_1$ of its own past predictions
and of the corresponding error signals $\ldots, \errsig_3, \errsig_2, \errsig_1$ obtained from the atomic reference,
it must make a prediction $\predfreq_0$ of the LO frequency in the next operating cycle.
The prediction should take into account the temporal correlations of the LO noise,
which dictate the timescale over which past measurement results remain relevant to predicting future LO behaviour.

We begin by considering the simple integrator,
the basic building block of the servo algorithm used in most contemporary optical frequency standards~\cite{Peik2006}.
In our notation,
the simple integrator makes the prediction
\begin{align}
  \predfreq_0 = \predfreq_1 + \gain \errsig_1
  \label{eqn:simpleinteform}
\end{align}
where $\gain$ is a dimensionless gain specifying the fraction of the frequency error measured in the last cycle to apply as a correction to the last prediction.
The prediction can also be expressed in terms of past estimates of the LO's fractional frequency deviation $\{\estfreq_k\}$ as follows:
\begin{equation}
  \predfreq_0 = \sum_{k=1}^{\infty} \gain (1-\gain)^{k-1} \estfreq_k.
  \label{eqn:simpleintyform}
\end{equation}
While Eqn.~\ref{eqn:simpleinteform} is easier to implement,
Eqn.~\ref{eqn:simpleintyform} is easier to reason about because the statistical properties of the estimated LO frequency $\estfreq$
are mostly determined by the LO noise and by the measurement noise of the reference,
depending only weakly on the design of the servo controller itself.
To a good first approximation, then, we can take the fluctuations and correlations of the $\{\estfreq_k\}$ as given,
and try to choose $\gain$ so as to minimise the error of the prediction $\predfreq_0$.

It is instructive to study the broader class of linear predictors,
whose predictions are weighted averages of past LO frequency estimates of the form
\begin{equation}
  \predfreq_0 = \sum_{k} \weight_k \estfreq_k,
  \label{eqn:defineglc}
\end{equation}
where the weights $\weight_k$ are required to satisfy the normalisation condition
\begin{equation}
  \sum_k \weight_k = 1.
  \label{eqn:weightsnorm}
\end{equation}
The simple integrator of Eqn.~\ref{eqn:simpleintyform} is a special case of a linear predictor,
with $\weight_k = \gain (1-\gain)^{k-1}$.
The optimisation of such linear predictors has been studied extensively since the pioneering work of Wiener~\cite{Wiener1942} and Kolmogorov~\cite{Kolmogorov1941} (see e.g. Ref.~\cite{Vaidyanathan2008}).
Here we derive the minimum-mean-squared-error predictor in a form similar to that used for ordinary kriging in geostatistics~(see e.g. \cite{Isaaks1989}).
We begin by computing the mean squared difference between the prediction $\predfreq_0$ and the next frequency \emph{estimate} $\estfreq_0$:
\begin{align}
  \avg{(\predfreq_0 - \estfreq_0)^2} &= \avg*{\sum_j \sum_k \weight_j \weight_k (\estfreq_j - \estfreq_0) (\estfreq_k - \estfreq_0)} \\
  &= \tpose{\vec{\weight}}\vec{\corr}\vec{\weight}, \label{eqn:predvar}
\end{align}
where we have collected the weights $\{\weight_k\}$ into a vector $\vec{\weight}$ and introduced the two-sample covariance matrix for the estimated LO frequency,
whose entries are defined as 
\begin{equation}
  \corr_{jk} = \avg{(\estfreq_j - \estfreq_0)(\estfreq_k - \estfreq_0)}.
  \label{eqn:defineC}
\end{equation}
Note that $\avg{(\predfreq_0 - \estfreq_0)^2}$ is not the same as
the mean squared prediction error $\avg{(\predfreq_0 - \lofreq_0)^2}$,
since it also includes the noise of the atomic reference which estimates that error.
Provided, however, that the atomic reference is unbiased,
the same choice of weights will minimise either measure of noise,
so we proceed to minimise Eqn.~\ref{eqn:predvar} and find that the optimum weights satisfy
\begin{equation}
  \vec{\corr} \vec{\weight} = \lambda \begin{pmatrix} 1 \\ 1 \\ 1 \\ 1 \\ \vdots \end{pmatrix} \label{eqn:weightlineq}
\end{equation}
with $\lambda$ a Lagrange multiplier that must be chosen to satisfy the normalisation constraint of Eqn.~\ref{eqn:weightsnorm}.
Thus the optimal weights can be found by solving Eqn.~\ref{eqn:weightlineq} for $\vec\weight / \lambda$
and normalising the result,
provided that one knows the covariance matrix $\vec\corr$.
If the noise properties of the components in the frequency standard are known,
then $\vec\corr$ can be computed simply as the sum of matrices for each independent noise process.
Explicit expressions for the $\vec\corr$ matrix associated with a known noise spectrum are provided in Appendix~\ref{sec:explicitC}.
As discussed in Sec.~\ref{sec:online},
$\vec\corr$ can also be estimated, and the servo controller optimised,
\emph{without} prior knowledge of the system noise properties,
using only data generated during normal clock operation.

Although linear predictors with arbitrary coefficients
are not difficult to implement following Eqn.~\ref{eqn:defineglc},
one can also use the preceding formalism to optimise the gain of conventional integrators.
Appendix~\ref{sec:optgain} derives an explicit, albeit cumbersome, formula for the optimal integrator gain given known noise model parameters.
Alternatively, one can use Eqn.~\ref{eqn:weightlineq} to choose a vector of weights for a hypothetical linear predictor
and then simply set the integrator gain to the leading entry of this vector $\gain = \weight_1$.
Simulations show that for common power-law noise processes,
the resulting integrating servo performs almost as well as the optimal linear predictor,
with a penalty of less than \SI{10}{\percent} in the prediction variance.
The formalism we have developed can thus be used to optimise the parameters of a conventional integrating servo algorithm,
without requiring any modifications to an already-running clock experiment.
As we will see, this optimisation can be performed even without prior knowledge of the experiment's noise characteristics.

\begin{figure}
  \centering
  \includegraphics{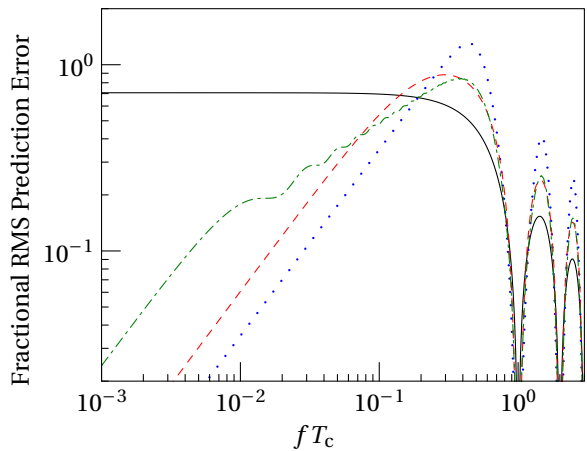}
  \caption{\CHANGED{Servo response spectra,
    expressed as the RMS prediction error
    caused by a unit-amplitude frequency modulation of the LO at frequency $f$,
    assuming a noiseless reference.
    Black solid line: no servo (open-loop).
    Dotted blue line: optimal linear predictor for pure random-walk noise (integrator of gain $\gain=1.27$).
    Chain-dotted green line: linear predictor optimised for pure flicker-frequency noise,
    with a 50-cycle memory.
    Dashed red line: best integrator for pure flicker-frequency noise ($\gain = 0.7$).
     }
  } \label{fig:servospectra}
\end{figure}

\CHANGED{
It may be helpful to visualise the spectral response of optimised servos.
Fig.~\ref{fig:servospectra} shows,
for a few simple cases,
the RMS magnitude of the prediction error caused by a frequency modulation of the LO at some frequency $f$.
In the absence of servo correction (black solid line), the response is flat at low frequencies but falls off as $1/f$ at high frequencies due to the averaging of the LO frequency within each interrogation cycle.
Note that the servo prediction error vanishes at those frequencies to which the reference is insensitive:
noise at these frequencies cannot be removed from the clock output,
but it does not disturb the atomic reference and is therefore irrelevant as far as the servo is concerned.
For a white-noise dominated system,
the optimal controller has the lowest practical gain,
or equivalently averages as much history as is available.
The optimal spectral response function in this case looks essentially identical to the black line.
For pure random-walk noise, which falls off rapidly at high frequencies,
it is worth increasing the sensitivity to high-frequency noise in order to obtain better suppression at low frequencies: the optimal controller is then an integrator with a gain $\gain=1.27$
(blue dotted line)\footnote{%
  Note that a random-walk-dominated system can require an integrator gain greater than unity.
  If, for instance, the last measured frequency was higher than the corresponding prediction,
  it is likely that the LO frequency was random-walking upwards during the last measurement,
  and thus that the frequency at the end of the measurement was higher than the average frequency   during the measurement.
  The apparent over-correction implied by $\gain>1$ is accounting for this difference.
  The integrating servo is still stable as long as $\gain<2$.}.
The power spectral density of random-walk noise $\propto f^{-2}$ combines with the $\propto f^2$ (power) response of an integrator to yield a flat spectrum of contributions to the prediction error.
In the intermediate case of flicker noise,
the same flat spectrum could be achieved by a controller with a power response $\propto f$,
i.e. an amplitude response $\propto \sqrt{f}$.
The optimal 50-term linear controller (green chain-dotted line) approximates this behaviour in the range of frequencies it can observe,
from roughly $1 / (100 \Tcyc)$ up to $1 / (2 \Tcyc)$.
At very low frequencies, corresponding to fluctuations slower than the 50-cycle memory of the controller, the response falls back to that of an integrator.
A simple integrator cannot have a $\sqrt{f}$ amplitude response,
so the best integrator for pure flicker noise (red dotted line) is more sensitive to fluctuations with periods of a few cycles.
As a result it performs about \SI{5}{\percent} worse than the more general linear predictor.
}

\begin{figure}
  \centering
  \includegraphics{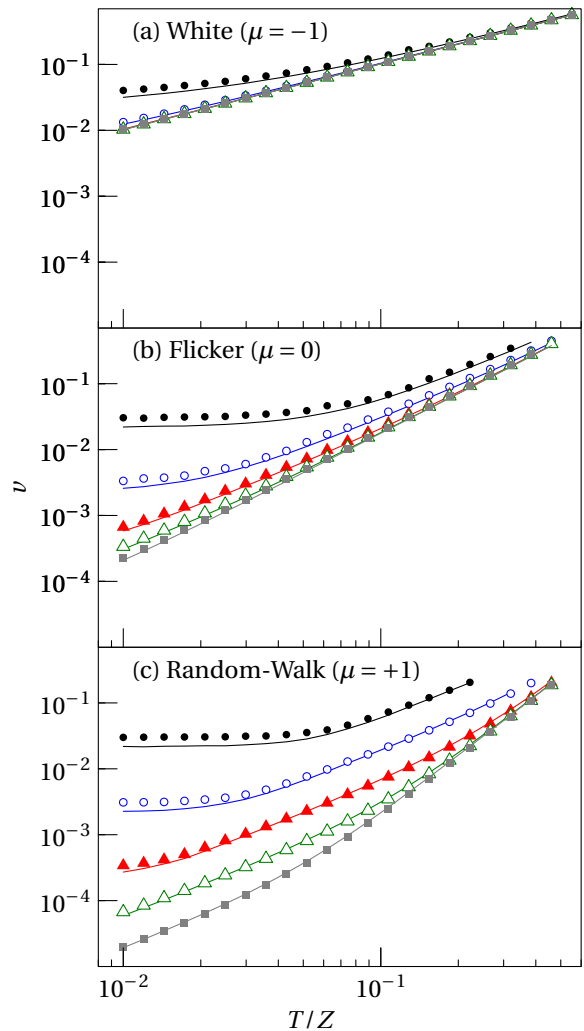}
  \caption{
    Variance of servo prediction errors,
    expressed as the variance $v$ of the phase accumulated in a Ramsey interrogation,
    as a function of probe time
    for a white-noise (top) flicker- (middle) or random-walk-limited (bottom) LO.
    The simulated clocks \CHANGED{run for \num{2e6} Ramsey interrogations} of
    1 (solid black circles),
    10 (open blue circles),
    \num{e2} (solid red triangles),
    \num{e3} (open green triangles)
    or \num{e4} (solid grey squares) uncorrelated atoms.
    Symbols show the performance of integrating controllers optimised without knowledge of the LO noise as discussed in Sec.~\ref{sec:online}.
    Solid lines show the performance of linear predictors using the last $\Csize=50$ frequency estimates,
    designed with knowledge of the LO noise properties as discussed in Sec.~\ref{sec:ctrldesign}.
  } \label{fig:predvar}
\end{figure}

\CHANGED{To gauge their impact on clock performance,}
we quantify the scale of the servo's prediction errors by the dimensionless variance
\begin{equation}
  \priorvar=\avg*{\left[\left(\lofreq - \predfreq\right)\wnom\Tprobe\right]^2}
  \label{eqn:definepriorvar}
\end{equation}
of the phase accumulated in the Ramsey interrogation of duration $\Tprobe$.
The variance $\priorvar$ plays an important role in determining both the robustness of the lock to atomic resonance
and the achievable long-term stability (see Sec.~\ref{sec:ltstab}),
so that it is worth studying its behaviour.
The solid lines in Fig.~\ref{fig:predvar} illustrate the performance of linear predictors,
based on simulations of clock operation with between 1 and \num{e4} atoms in the reference.
As a function of the choice of probe time $\Tprobe/\Tnought$,
and thus of the ratio between LO noise and measurement noise,
one can distinguish three qualitatively different regimes.
In the limit of large atom numbers and long probe times,
the simulations approach an $N$-independent limit
\begin{equation}
  \priorvar = \varprefactor \left(\frac{\Tprobe}{\Tnought}\right)^{2+\noisetype}.
  \label{eqn:highNv}
\end{equation}
This is the scaling one expects for the case where
the LO noise completely dominates the measurement noise of the reference,
given the postulated power-law scaling of the LO noise with exponent $\noisetype$ (c.f. Sec.~\ref{sec:notation})
and the definition of $\Tnought$ in Eqn.~\ref{eqn:defineTnought}.
The proportionality constant $\varprefactor$ can be estimated
from the simulations or derived using the formulae in Appendix~\ref{sec:explicitC},
and varies between 1
(for a white-noise-dominated LO)
and 2 (for a random-walk-noise-dominated LO).
In the opposite limit of low atom number or short probe time,
the (white) quantum projection noise dominates
and the servo performance depends only on the number of measurements $\Csize$ which it averages in making its prediction,
and thus 
\begin{equation}
  \priorvar \approx \frac{1}{N\Csize}
  \label{eqn:lowNv}
\end{equation}
for sufficiently short probe times.
For $\Csize \gtrsim 20$ this limit has no impact on correctly optimised clock operation (see Sec.~\ref{sec:refparams}).
Between the two limits considered above
there is a trade-off between averaging many measurements to reduce the impact of measurement noise
and considering only the recent measurements most relevant to the LO's current frequency.
We know of no simple, accurate expression for the achievable servo performance in this intermediate regime,
but the rough scaling that one would expect from the aforementioned trade-off
\begin{equation}
  \priorvar \propto \frac{1}{\sqrt{N}} \left(\frac{\Tprobe}{\Tnought}\right)^{1+\noisetype/2}
  \label{eqn:midNv}
\end{equation}
does hold in simulations.

So far, we have discussed the simple integrator and its generalisations.
Practical frequency standards, however, must use a double-integrator to correct for steady drifts in the LO frequency~\cite{Peik2006}.
With the addition of the second integrator,
the servo predictions become
\begin{align}
  \predfreq_0 =& \predfreq_1 + \gain \errsig_1 + \driftgain\sum_k \errsig_k \\
\intertext{or}
  =& \sum_{k} \weight_k \estfreq_k + \driftgain\sum_k \errsig_k.
  \label{eqn:definedblint}
\end{align}
Aside from its role in suppressing steady-state frequency errors with drifting LOs,
the second integrator is necessary for the servo to have enough low-frequency gain to attain the projection noise limit in many-atom clocks, a point to which we will return in Sec.~\ref{sec:dblint}.
However, as long as its gain $\driftgain$ is chosen low enough to avoid servo oscillations,
the additional integrator has only a negligible impact on the variance of the prediction errors%
\footnote{
  This is best understood by considering the action of the servo in the frequency domain:
  the variance of the prediction errors depends on the feedback gain at a frequency corresponding to the clock cycle rate,
  whereas the second integrator only contributes feedback gain at much lower frequencies.
  The effect of the second servo is visible in the correlations between prediction errors, not in their variance.
}.
The controller can thus be designed by optimising a simple integrator or linear predictor as discussed above,
and then adding the drift-correction integrator with a gain $\driftgain \ll \weight_1 = \gain$.

Besides minimising prediction variance,
another desirable feature in practical servo controllers is \emph{robustness},
the quality of remaining locked to the (correct) atomic resonance for long periods.
In principle the two qualities are distinct,
but we find empirically that for well-optimised servos they are tightly coupled.
In simulations of clocks with a wide range of atom numbers (i.e. reference signal-to-noise ratios),
we find that the rate at which a clock hops to different Ramsey fringes depends,
for a given LO and a fully optimised servo,
only on the prediction variance $\priorvar$.
Suboptimal servos (such as integrators with incorrectly chosen gain)
have both greater prediction variance $\priorvar$ and a higher rate of fringe hops for a given $\priorvar$,
so that they are less robust as well as noisier.
We conjecture that the best servos are simultaneously the most robust and the least noisy,
so that there is no need to choose between the two qualities provided that one can, in fact, find this optimal servo design.

\section{On-Line Servo Optimisation and Noise Characterisation}
\label{sec:online}

In practice, the noise spectrum of the LO may not be known accurately.
A significant benefit of the formalism presented in the previous section
is that it allows one to optimise the servo controller without prior knowledge of the LO noise properties.
This is possible because the definition of $\vec\corr$ in Eqn.~\ref{eqn:defineC}
involves only the \emph{estimated} LO frequency error in each clock cycle,
which is routinely recorded in normal clock operation%
\footnote{
  In some implementations, the estimated frequency might not be recorded as such,
  but it can be obtained by adding the servo predictions to the frequency error signal reported by the atomic reference.
}.
As a demonstration of such optimisation,
we have run clock simulations with integrating servos whose gains were chosen,
without knowledge of the true LO noise,
by the following empirical procedure:
\begin{enumerate}
\item Start by setting the gain to $\gain=0.2$,
  an arbitrary but reasonable initial value chosen to allow reliable, if suboptimal, clock operation under a wide range of conditions.
\item Simulate the clock for \num{e4} cycles, corresponding to a few hours of operation for a typical contemporary frequency standard.
\item Compute $\vec\corr$ according to Eqn.~\ref{eqn:defineC} and thence the vector $\vec\weight$ of optimal weights.
  Set $\gain = \weight_1$
  \footnote{
    In cases where the noise is nearly white,
    this can lead to a theoretically optimal but practically useless controller with near-zero gain.
    To avoid this problem, we impose a minimum gain of \num{0.04} to ensure a finite servo attack time.
    This bound is low enough that it does not affect the achievable clock stability.
  }.
\item Simulate the clock with the newly optimised servo and reoptimise,
  repeating as necessary.
\end{enumerate}
Even when the servo gain is initially chosen blindly,
we find that five rounds of optimisation suffice for the gain $\gain$ to converge to a value that yields performance indistinguishable from that of an integrator designed with full knowledge of the LO noise spectrum.
Under more realistic conditions,
where the initial choice of servo parameters reflects some prior knowledge of the LO performance,
the optimisation could be performed much more quickly.
The symbols in Fig.~\ref{fig:predvar} show the prediction variance of such empirically-optimised integrators,
which can be compared to the performance of optimal linear predictors shown as solid lines.
For clocks operated near their optimal probe times (to be discussed in Sec.~\ref{sec:ltstab}),
the difference in prediction variance is less than \SI{10}{\percent}.
Thus, it is possible to develop controllers that take full advantage of the time correlations in the LO noise even without independent knowledge of those correlations.

Unfortunately, it is not always possible to verify the servo performance directly,
because the observed variance of the error signal $\errsig$ contains contributions both from the servo prediction error and from measurement noise of the reference.
For a single-atom clock this problem is insurmountable:
the observed fluctuations of a binary error signal must correspond to quantum projection noise,
independent of servo performance.
For a many-atom clock where the detection noise is well-characterised
it is possible to measure the servo prediction variance as an increase in the fluctuations of the error signal,
but the resulting estimates are generally optimistic.
As discussed in Sec.~\ref{sec:ltstab},
even in a correctly optimised clock there will be unavoidable ambiguities in interpreting the error signal (e.g. $2\pi$ phase slips in Ramsey interrogation) and the resulting measurement errors contribute to the servo prediction variance without being observable in the experimentally recorded measurement data.

\begin{figure}
  \centering
  \includegraphics{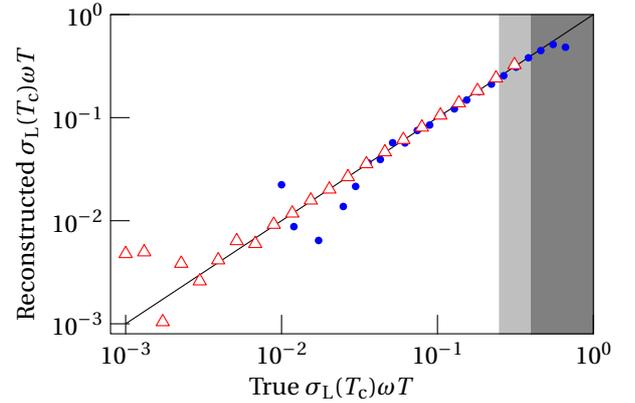}
  \caption{
    Normalised Allan deviation of flicker (blue circles) or random-walk (red triangles) LO noise
    reconstructed from correlations in the error signal of a single-atom clock
    as a function of the actual noise level.
    Black line marks correct reconstruction.
    Light and dark grey regions mark the noise levels at which the clock servo hops between Ramsey fringes for
    random walk and flicker noise respectively.
  } \label{fig:reconstructlo}
\end{figure}

One can, however, use the correlation matrix $\vec\corr$ estimated during clock operation to partially characterise the LO.
Although white noise of the LO is indistinguishable from measurement noise in the atomic reference,
flicker or random-walk noise can produce detectable temporal correlations even when their contributions to the total measurement variance are small.
By fitting the estimated $\vec\corr$ to a linear combination
\begin{equation}
  \vec\corr =  \wndev^2\of\Tcyc \vec\wncorr + \ffdev^2 \vec\ffcorr + \rwdev^2\of\Tcyc \vec\rwcorr \label{eqn:lorecon}
\end{equation}
of the correlation matrices expected for white, flicker and random-walk noise,
one can obtain estimates of the Allan variances associated with each class of noise process,
which are simply the coefficients in the linear combination.
Explicit expressions for the correlation matrices associated with arbitrary noise spectra are provided in Appendix.~\ref{sec:explicitC}.
Fig.~\ref{fig:reconstructlo}, for instance,
shows the flicker and random-walk Allan deviations reconstructed in this fashion from a correlation matrix $\vec\corr$ estimated from \num{2e6} cycles of operation of a simulated single-atom clock,
normalised to quantum projection noise,
as a function of the true level of the respective noise processes.
Random-walk noise,
whose correlations differ more strongly from those of the white measurement noise,
is easier to detect,
but as seen in Fig.~\ref{fig:reconstructlo} both flicker and random walk noise can be reliably estimated from levels too low to affect the clock's instability (variance less than \SI{1}{\percent} of projection noise)
up to levels that would be unacceptably high in normal operation,
when the clock servo is jumping between Ramsey fringes.
Although this method provides much less detailed information on the LO noise spectrum than does the optical spectrum analyser of Ref.~\cite{Bishof2013},
it requires no measurements beyond those performed as part of the clock's normal operation.
It can therefore be used to monitor the LO while the clock is running,
even in a single-ion frequency standard,
providing an early warning of performance degradations as well as information useful for the optimisation of interrogation parameters in the atomic reference (see Sec.~\ref{sec:refparams}).

\section{Impact of servo performance on long-term stability}
\label{sec:ltstab}

Any measurement performed on a finite number of atoms can yield only a finite number of possible results.
The optimisation of the measurement protocol thus involves a compromise
between fine resolution over a narrow usable domain of LO frequencies and
coarse frequency resolution over a broader domain.
The best compromise depends on the range of frequencies which might plausibly have to be measured by the reference,
i.e. on the variance of servo prediction errors.
In this section we study this compromise,
showing that it leads to a finite optimal probe time
and an overall limit on the long-term stability of clocks with noisy LOs.

\begin{figure}
  \centering
  \includegraphics{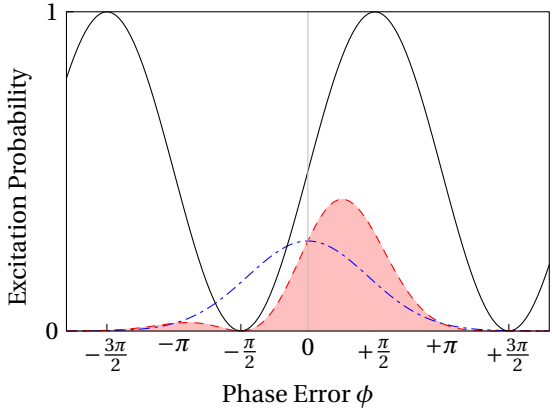}
  \caption{
    Schematic illustration of information gain in a single cycle of clock operation.
    Blue chain-dotted Gaussian: Prior knowledge of the LO frequency.
    Black solid sinusoid: Ramsey excitation spectrum.
    Red dashed line: Posterior knowledge of the LO frequency given that an atom was detected in the excited state.
    If no excitation is detected, then the posterior distribution is the mirror image about $\phi=0$ of the one shown here.
  } \label{fig:priorpost}
\end{figure}

By way of illustration, consider Ramsey interrogation of a single atom,
where the probability to find the atom in the excited state depends on the phase error
$\phi = (\lofreq - \predfreq)\wnom\Tprobe$ as
\begin{equation}
  \Pexc = \frac{1 + \sin \phi}{2}.
  \label{eqn:ramseyfringe}
\end{equation}
This excitation spectrum is shown as a solid line of Fig.~\ref{fig:priorpost}.
Let us assume that,
before any measurement,
our best knowledge of the (corrected) LO frequency
is represented by the chain-dotted distribution,
corresponding to the distribution of LO prediction errors.
Our best knowledge after a measurement on a single atom,
in the event that it is found in the excited state,
is shown by the dashed probability distribution.
In this case we can be certain that the phase was not near $\phi=-\pi/2$
(since the excitation probability would then have been 0),
and we can be reasonably confident that it lies in the region between 0 and $\pi$,
but we cannot rule out that it lies near $-\pi$,
where both the excitation probability and the prior probability distribution are non-negligible.
Varying the probe time,
and hence the spacing of the Ramsey fringes,
involves a trade-off:
increasing $\Tprobe$ narrows the main lobe of the posterior distribution,
thanks to the steeper slope of the excitation signal,
but also increases the weight of secondary lobes due to other fringes of the Ramsey spectrum.

To quantify this trade-off more formally, we consider Ramsey interrogation of $N$ uncorrelated atoms,
taking the prior distribution (chain-dotted curve in Fig.~\ref{fig:priorpost}) to be a Gaussian of variance $\priorvar$
\begin{equation}
  \prior\of{\phi} = \frac{\ee^{-\frac{\phi^2}{2 \priorvar}}}{\sqrt{2\pi \priorvar}}.
  \label{eqn:prior}
\end{equation}
This distribution encodes, formally, all that is known about the LO frequency before the atomic measurement result becomes available.
In a practical sense, it is the distribution of servo prediction errors:
if the servo prediction were perfect ($\predfreq = \lofreq$) then the phase accumulated in the Ramsey interrogation would be zero.
\CHANGED{
The ansatz of Eqn.~\ref{eqn:prior} thus amounts to an assumption that the servo prediction errors are normally distributed.
Although our simulations show some small deviations from the normal distribution,
amounting to a negative excess kurtosis of a few percent with a single-atom reference that produces a binary error signal,
the Gaussian ansatz is a surprisingly good approximation.
As we will see, it leads to simple analytical results which agree well with more detailed simulations.}

The reference does not, unfortunately, supply us with the expectation value $\Pexc$.
Rather, the measurement yields a random fraction $\excfrac$ of atoms detected in the excited state 
that fluctuates about the expectation value $\Pexc$ due to measurement noise of the atomic reference.
In the absence of technical noise on the reference signal, the variance of $\excfrac$ is
\begin{align}
  \var\of\excfrac =& \avg*{(\Pexc - \avg{\Pexc})^2 + \frac{\Pexc (1 - \Pexc)}{N}} \\
  =& \frac{\ee^{-v}\sinh{v}}{4} + \frac{1 - \ee^{-v}\sinh{v}}{4N}
  \label{eqn:excvar}
\end{align}
where the averages $\avg{\cdot}$ are taken over the prior distribution $\prior\of{\phi}$.
The first term expresses the fluctuations in the measured excitation fraction due to actual changes in the $\phi$-dependent excitation probability,
while the second corresponds to quantum projection noise of the binomially-distributed excitation signal.
The usefulness of the excitation fraction in estimating the frequency depends on
the covariance of the two quantities
\begin{equation}
  \cov\of{\phi, \excfrac} = \avg*{\phi \excfrac} - \avg{\phi}\avg{\excfrac} = \frac{v}{2}\ee^{-\frac{v}{2}}
  \label{eqn:excphicorr}
\end{equation}
which determines how much weight should be given to $\excfrac$
in constructing the posterior estimate of the LO frequency.
Choosing the weight to minimise the variance $\postvar$ of the error in this posterior estimate,
we find (c.f. Appendix~\ref{sec:postvar}):
\begin{align}
  \postvar &= \priorvar - \frac{\cov\of{\phi, \excfrac}^2}{\var\of\excfrac} \label{eqn:linearvarred} \\
  &= \priorvar \left[1 - \frac{N\priorvar}{(N - 1)\sinh \priorvar + \ee^\priorvar}\right].
  \label{eqn:postvar}
\end{align}
This posterior variance combines information from the measurement with information that was known beforehand.
In order to isolate the contribution of the former,
we define an effective measurement variance $\msmtvar$ by
\begin{align}
  \frac{1}{\postvar} = \frac{1}{\priorvar} + \frac{1}{\msmtvar}.
  \label{eqn:msmtvardef}
\end{align}
This is the usual relation for the variance of a (posterior) estimate obtained by an optimal linear combination of two \emph{independent} pieces of information.
$\msmtvar$ is thus the variance of a hypothetical measurement,
one that could be interpreted without any prior knowledge,
and which would reduce our uncertainty on the LO frequency as much as did the actual measurement.
For the case we consider,
\begin{align}
  \msmtvar &= \frac{\ee^\priorvar}{N} + \left(1 - \frac{1}{N}\right) \sinh \priorvar - \priorvar \label{eqn:msmtvar}\\
  &\approx \frac{1}{N} + \frac{\priorvar^2}{2 N} + \frac{\priorvar^3}{6} + \cdots \label{eqn:msmtvarapprox}
\end{align}
where, in the second line, we have expanded the effective measurement variance in powers of the prior variance.
The first term is the conventional QPN on the measurement of the phase,
valid when $\priorvar$ is small and the corrected LO frequency is known
\emph{a priori} to be well centred on the Ramsey fringe.
The third term reflects the additional uncertainty arising
when the corrected LO frequency can lie outside the range where the reference produces a meaningful result.
This term is independent of atom number,
and dominates the effective measurement variance as $\priorvar$ approaches 1.

\begin{figure}
  \centering
  \includegraphics{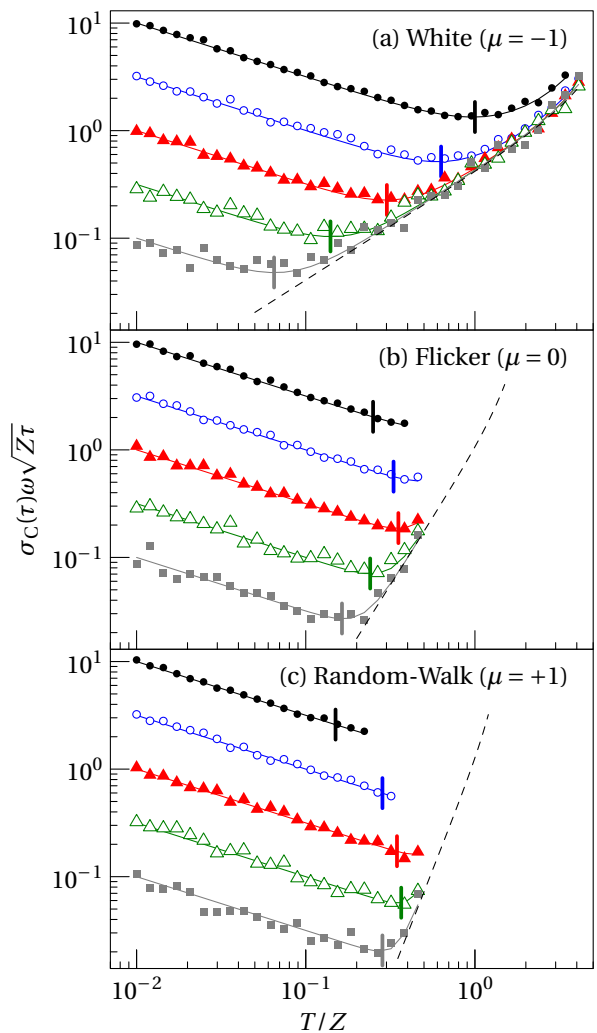}
  \caption{
    Long-term instability as a function of probe time for clocks using
    Ramsey interrogation of
    1 (solid black circles),
    10 (open blue circles),
    \num{e2} (solid red triangles),
    \num{e3} (open green triangles)
    or \num{e4} (solid grey squares) uncorrelated atoms.
    Solid lines show the prediction of Eqn.~\ref{eqn:modsql}, without free parameters.
    Vertical bars mark the recommended interrogation time given in Table~\ref{tbl:Topt}.
    Dashed line marks the limit for perfect phase estimation with no projection noise. 
    The three graphs are, from top to bottom, for a white-, flicker-, or random-walk-dominated LO.
    \CHANGED{Simulations ran for \num{2e6} clock cycles.}
  } \label{fig:ltstab}
\end{figure}

Replacing the standard phase variance of Eqn.~\ref{eqn:defineqpn} by the effective measurement variance in Eqn.~\ref{eqn:definesql} yields a new prediction for clock stability in the limit of large averaging time $\tau$,
one that accounts for the effects of limited prior information in each interrogation:
\begin{equation}
  \clkadev\of\tau \rightarrow \frac{1}{\wnom\Tprobe}\sqrt{\frac{\Tcyc}{\tau}}
  \sqrt{\frac{\ee^\priorvar}{N} + \left(1 - \frac{1}{N}\right) \sinh \priorvar - \priorvar}. \label{eqn:modsql}
\end{equation}
\CHANGED{To demonstrate the validity of the simplifying approximations made in our model,
such as the Gaussian ansatz for the distribution of servo errors,
we compare the zero-free-parameter prediction of Eqn.~\ref{eqn:modsql}
(Fig.~\ref{fig:ltstab}, solid lines)
to the stability of clocks simulated without making those approximations (Fig.~\ref{fig:ltstab}, symbols).
Clocks with white-, flicker-, and random-walk-noise-limited LOs were simulated
using Ramsey interrogation of uncorrelated atoms with no dead time ($\Tcyc=\Tprobe$).
Each point in Fig.~\ref{fig:ltstab}} is obtained from a simulation of \num{2e6} cycles of clock operation.
The Allan deviation is computed for a time $\tau$
long enough that the instability has reached the asymptotic $1/\sqrt{\tau}$ regime
(corresponding to \num{2e4} cycles of clock operation),
then rescaled to a fixed averaging time $\Tnought$
and normalised to a fixed noise level $\loadev\of\Tnought$
to obtain a dimensionless result that is comparable across systems.
The graphs thus show the achievable long-term instability as a function of the choice of probe time.

When the probe time is short, the Ramsey fringe is broad and $\priorvar$ is small,
the instability improves as $1/\sqrt{\Tprobe}$ as conventionally expected.
The improvement with increasing probe time stops either when the additional $\priorvar$-dependent terms in Eqn.~\ref{eqn:modsql} grow important or when the servo can no longer reliably lock the LO to the reference transition:
the curves in Fig.~\ref{fig:ltstab} end when the fringe-hop rate reaches 1 per 2 million cycles.
As $N$ increases, quantum projection noise is reduced relative to LO noise
and it becomes advantageous to reduce the probe time so as to be less sensitive to the latter.
Thus, the optimal probe time gets shorter with increasing atom number,
and the fully optimised clock instability does not scale as $N^{-1/2}$.
The asymptotic scaling with $N$ is given in Table~\ref{tbl:Nscaling}.
For white noise, the most extreme case,
the optimal probe time scales as $N^{-1/3}$ in the large-$N$ limit,
leading to a $N^{-1/3}$ scaling of the long-term Allan deviation.
For flicker or random-walk noise,
$\priorvar$ falls off more steeply as the probe time is shortened (see Eqn.~\ref{eqn:highNv}),
so that the the optimal probe time is less sensitive to atom number
and a scaling closer to the conventional QPN limit is obtained.
In the absence of projection noise,
i.e. in the limit $N\rightarrow\infty$,
the servo performance limit of Eqn.~\ref{eqn:highNv} combines with Eqn.~\ref{eqn:modsql} to yield a general measurement-noise-independent limit on clock instability, which we plot as dashed lines in Fig.~\ref{fig:ltstab}.
This limit arises solely from the unpredictability of the LO noise and from the finite domain over which the Ramsey error signal can be unambiguously interpreted.

\begin{table}
  \begin{tabular}{ll}
    \hline\noalign{\smallskip}
    LO Noise Type & Asymptotic Scaling\\
      & of $\clkadev\of\tau\wnom\sqrt{\Tnought\tau}$ \\
    \noalign{\smallskip}\hline\noalign{\smallskip}
    White & $\propto N^{-1/3}$ \\
    Flicker & $\propto N^{-5/12}$ \\
    Rnd. Walk & $\propto N^{-4/9}$ \\
    \noalign{\smallskip}\hline
  \end{tabular}
  \caption{
    Asymptotic scaling of LO-limited clock instability with atom number.
    The scaling differs from the conventional $N^{-1/2}$ QPN limit
    because the optimum probe time decreases with increasing atom number.
   } \label{tbl:Nscaling}
 \end{table}

Strictly speaking, Eqn.~\ref{eqn:linearvarred} holds
only if the estimated frequency error is a linear function of the measured excitation fraction.
Since the excitation probability is a non-linear (e.g. sinusoidal) function of the LO frequency error,
one might hope to do better than the estimated performance of Eqn.~\ref{eqn:modsql} by using a non-linear function
to convert the excitation fraction to a frequency error estimate.
Simulations show, however, that correcting for the curvature of the Ramsey fringe by estimating the accumulated phase as $\arcsin(2\excfrac-1)$ rather than simply $2\excfrac - 1$ has no significant effect on $\priorvar$ or on the achievable long-term clock stability.
One can understand this finding by noting that,
when $\priorvar$ is large enough that the curvature within a single Ramsey fringe is significant,
the effect of unavoidable ambiguities such as the secondary lobe in Fig.~\ref{fig:priorpost} is much larger and dominates the posterior variance.

\section{Guidelines for Interrogation Parameters}
\label{sec:refparams}

\begin{table}
  \begin{tabular}{lcc}
    \hline\noalign{\smallskip}
    LO Noise & Safe &  Asymptotic Optimum \\
      & $\Tprobe/\Tnought$ & $\Tprobe/\Tnought$ \\
    \noalign{\smallskip}\hline\noalign{\smallskip}
    White & -- & $\min\of*{N^{-1/5},\,1.4\,N^{-1/3}}$ \\
    Flicker & $0.4 - 0.15\,N^{-1/3}$ & $0.76\,N^{-1/6}$ \\
    Rnd. Walk & $0.4-0.25\,N^{-1/3}$ & $0.79\,N^{-1/9}$ \\
    \noalign{\smallskip}\hline
  \end{tabular}
  \caption{
    Recommendations for the choice of Ramsey interrogation time.
    The last column gives the optimal probe time in the limit of many atoms.
    There is nothing to be gained by probing longer than this time.
    It may be necessary to use shorter probe times to avoid fringe hops;
    a suggested safe upper bound on the probe time is given in the second column.
   } \label{tbl:Topt}
 \end{table}
 
To choose the operating parameters for a clock,
one can in general use the formalism of Sec.~\ref{sec:ctrldesign} to predict the servo error variance $v$ as a function of those operating parameters
and Eqn.~\ref{eqn:modsql} to predict the resulting long-term stability,
which can then be optimised.
Table~\ref{tbl:Topt}, for example, provides recommended Ramsey interrogation times
for clocks dominated by different types of power-law LO noise,
expressed as multiples of $\Tnought$.
In the many-atom limit,
the servo prediction errors become independent of the quantum projection noise
and we can solve Eqn.~\ref{eqn:modsql} to obtain the asymptotically optimal probe time
(last column of Table~\ref{tbl:Topt}).
Increasing the probe time beyond this optimum always leads to an increase in
effective measurement variance and long-term instability,
and is of no practical interest.
At small atom numbers,
shorter probe times are required to keep the servo controller robust against fringe hops.
The purely phenomenological bound in the second column of Table~\ref{tbl:Topt}
is chosen to be slightly shorter than the time for which we observe fringe-hops at a rate of 1 per million simulated clock cycles, with a \SI{20}{\percent} safety margin.
Our choice of maximum acceptable fringe-hop rate,
corresponding to a requirement that the clock remain locked to the correct fringe for a few days,
is arbitrary,
but as the onset of fringe-hopping is extremely steep
(the fringe-hop rate in simulations increases by two to three orders of magnitude when the probe time is doubled),
the maximum safe probe time is only weakly dependent on this choice of threshold.
A full optimisation of all common probe protocols in the presence of realistic experimental imperfections is beyond the scope of this work,
but we expect qualitatively similar behaviour from Rabi or hyper-Ramsey probing,
with somewhat longer optimal probe times and slightly degraded instability due to the increased width of the observed atomic resonance in theses schemes.
Conversely, we expect that clocks with significant dead time in their operating cycle will need to use somewhat shorter probe times to compensate for the servo's inability to correct unobserved LO frequency fluctuations,
which will lead to a $\priorvar$ higher than in our dead-time-free simulations.

\section{Constraints on the benefits of entangled atomic references}
\label{sec:entanglement}

\begin{figure}
  \centering
  \includegraphics{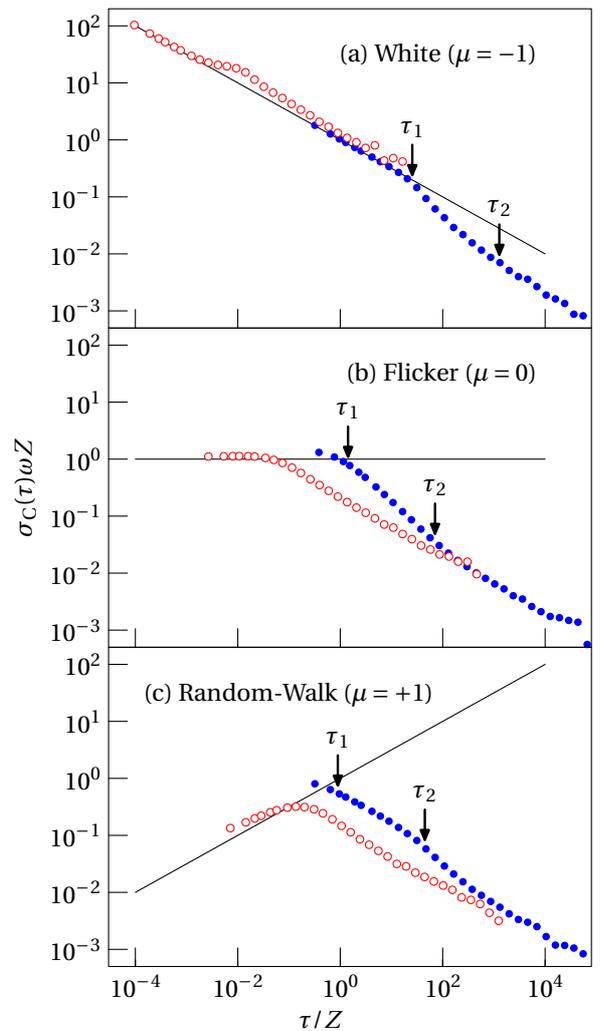}
  \caption{
    Simulated Allan deviation as a function of averaging time
    for optimised clocks using either 100 uncorrelated atoms (solid blue circles)
    or a maximally-correlated state of 100 atoms (open red circles).
    The latter must operate with shorter interrogation times,
    and thus the corresponding curves start earlier.
    The black line marks the Allan deviation of the free-running LO.
    The three graphs are, from top to bottom, for a white-, flicker-, or random-walk-dominated LO.
    Arrows mark the two integrator time constants for the case of uncorrelated atoms,
    as discussed in Sec.~\ref{sec:dblint}.
  } \label{fig:adev}
\end{figure}

The arguments developed in the preceding sections also apply to certain Ramsey-like protocols using entangled states in the atomic reference,
provided that LO noise is the limiting form of decoherence.
For instance, the scheme proposed in Ref.~\cite{Bollinger1996} and demonstrated in Refs.~\cite{Meyer2001,leibfried_creation_2005, Monz2011},
employing $N$ atoms in a maximally correlated state [$\ket{\psi} = \left(\ket{g}^{\otimes N} + \ket{e}^{\otimes N}\right)/\sqrt{2}$, where $\ket{g}$ and $\ket{e}$ are the atomic eigenstates],
is fully equivalent to Ramsey interrogation of a single atom with an $N$-fold enhanced transition frequency
by the corresponding harmonic of the LO radiation.
Now the long-term instability of a single-atom frequency standard can be expressed as
\begin{equation}
  \clkadev\of\tau = \frac{\stabconst}{\wnom \sqrt{\Tnought \tau}},
  \label{eqn:oneatomstab}
\end{equation}
with $\stabconst$ a dimensionless constant of order unity encoding the choice of probe time $\Tprobe / \Tnought$ and the additional contribution of LO noise at this probe time.
The long-term instability of the clock using maximally-correlated atoms thus becomes
\begin{equation}
  \clkadev\of\tau = \frac{\stabconst}{N\wnom \sqrt{\TnoughtGHZ \tau}},
  \label{eqn:GHZstab}
\end{equation}
where $\TnoughtGHZ$ is the noise timescale for the $N$-fold frequency-multiplied LO:
\begin{equation}
  \loadev\of{\TnoughtGHZcyc} N \wnom \TnoughtGHZ = \SI{1}{\radian}.
  \label{eqn:GHZtnought}
\end{equation}
In the dead-time-free limit $\TnoughtGHZcyc = \TnoughtGHZ$,
one can compare Eqn.~\ref{eqn:GHZtnought} with Eqn.~\ref{eqn:defineTnought} and find
\begin{equation}
  \frac{\TnoughtGHZ}{\Tnought} = N^\frac{-2}{2 + \noisetype},
  \label{eqn:GHZtnoughtscaling}
\end{equation}
with $\noisetype$ again describing the time-dependence of the LO Allan variance (see Sec.~\ref{sec:notation}).
The entangled clock's long-term instability thus scales as
\begin{equation}
  \clkadev\of\tau = \frac{\stabconst}{\wnom \sqrt{\Tnought \tau}} \frac{1}{N^\frac{1+\noisetype}{2+\noisetype}}.
  \label{eqn:GHZstabscaling}
\end{equation}
Thus, if the clock stability is limited by white LO noise ($\noisetype=-1$),
a reference using a maximally-correlated state of $N$ atoms performs no better than a reference using a \emph{single} atom,
and is in fact \emph{worse} than a reference using uncorrelated interrogation of the $N$ atoms. This scaling has been observed experimentally for correlated magnetic field noise in a 14~ion GHZ state~\cite{Monz2011}.
For flicker-floor LO noise,
the Allan deviation improves as $N^{-1/2}$ with $N$ maximally-entangled atoms,
very slightly better than the asymptotic $N^{-5/12}$ scaling achievable without entanglement,
but worse than the scaling achieved with unentangled atoms for $N<\num{e2}$.
It is only for random-walk LO noise that maximally entangled states offer measurable benefits,
with an $N^{-2/3}$ scaling of the long-term Allan deviation.
We illustrate these scalings in Fig.~\ref{fig:adev},
which plots the Allan deviation spectrum recorded in simulations of fully optimised clocks using either 100 uncorrelated atoms or a 100-atom maximally-correlated state for all three LO noise types.

Maximally entangled states reduce the signal-to-noise ratio of a measurement on $N$ atoms to that of a single qubit,
providing less new information per measurement but accelerating the clock cycle so that more measurements can be averaged.
That is why their use is advantageous with random-walk LO noise,
when fast measurements can take advantage of the reduced LO noise at short time scales.
Other approaches to the use of entanglement in atomic references,
such as spin squeezing~\cite{Kitagawa1993,Wineland1992,Wineland1994,Leroux2010:clock,Louchet-Chauvet2010,Hosten2016},
focus instead on improving the signal-to-noise ratio of the measurement,
and thus increasing the amount of new information obtained in each interrogation.
The error signal produced in such schemes has the same periodic ambiguities as in Ramsey interrogation of uncorrelated atoms,
and so they are subject to the same projection-noise-independent
$\sinh\priorvar - \priorvar \approx \priorvar^3/6$ limit on their effective measurement variance (c.f. Eqn.~\ref{eqn:msmtvarapprox}).
The additional noise introduced when servo prediction errors allow the anti-squeezed quadrature to contaminate the measurement result~\cite{Andre2004}
can in principle be eliminated by a suitable readout procedure~\cite{Borregard2013}
in which case we expect such interrogation protocols to offer benefits comparable to suppressing the projection noise by increasing atom number,
even with white- or flicker-floor limited LOs.

\section{Some Remarks on Short-Term Instability}
\label{sec:dblint}

So far we have focused on the long-term instability of the clock
once it reaches the asymptotic $\clkadev\propto1/\sqrt\tau$ regime,
without considering the averaging time required to reach this regime.
In general,
a clock reaches its asymptotic instability when the fluctuations
in the frequency of the output signal are dominated by the measurement noise of the atomic reference.
For single-ion clocks in which the signal-to-noise ratio of the atomic measurements is no better than 1,
this condition is reached at the servo attack time $\mainattack$,
as soon as the output signal of the clock stops following the free-running LO and is locked to the noisy signal from the atomic reference.
However, clocks using many atoms have a much higher signal-to-noise ratio,
i.e. the resolution of the error signal from their atomic reference is much finer than the LO frequency fluctuations that they can reliably measure.
In such clocks the optimal probe times are long enough that
the quantum projection noise is well below the LO-limited short-term instability.
In order to reach the asymptotic regime they must initially average down faster than $1/\sqrt{\tau}$.
This is possible provided that the servo has enough gain to suppress the measured LO frequency fluctuations.
A single integrator can suppress measured LO fluctuations by a factor $\sim\tau/\mainattack$ in the standard deviation%
\footnote{
  This is the time-domain equivalent of the observation that an integrating servo suppresses noise power at a frequency $f$
  by a factor proportional to $f^2$.
},
so that the clock instability initially averages down as $\tau^{-3/2}$ or $\tau^{-1}$ for white or flicker LO noise respectively.
This is fast enough to reach the measurement-noise limited regime in an averaging time of roughly $\sqrt{N}\mainattack$ or $N\mainattack$
for a white-noise-limited or flicker-floor limited LO respectively,
as seen in the first two graphs of Fig.~\ref{fig:adev}.
However, a single integrator can only suppress random-walk LO noise to a level that scales as $\tau^{-1/2}$,
which will not catch up with the measurement noise limit which is averaging down at the same rate.
A many-atom clock using only a single integrator would thus be
forever limited by the finite gain of the servo rather than by the noise of the atomic measurements.
It is only when a second integrator allows the servo to suppress noise by an additional factor of $\tau/\driftattack$ that the clock instability can average down as $\tau^{-3/2}$ until it reaches the measurement noise limit in a time $\sqrt{N}\driftattack$.
The third graph of Fig.~\ref{fig:adev} illustrates this behaviour,
with the uncorrelated 100-atom clock initially averaging down at the servo-limited rate of $\tau^{-1/2}$
until around $\driftattack\approx45\Tnought$.
The second integrator then allows the instability to catch up with the lower-lying asymptotic noise limit,
which it reaches around $\tau\approx400\Tnought$.
It is interesting to note that clocks using maximally-correlated states,
because they behave like single-atom clocks and are always measurement-noise limited,
would have an advantage in short-term instability even when their long-term instability is little better than that of a clock with uncorrelated atoms (lower two graphs of Fig.~\ref{fig:adev}).
This observation mirrors,
in a simpler setting,
the finding of Ref.~\cite{Kessler2014}.

Thus the second integrator in a clock servo,
beyond its role in correcting for linear drifts,
is also needed to suppress random-walk noise of the LO in many-atom clocks.
It is desirable to set the gain $\driftgain$ of this drift-correction integrator as high as possible,
in order to reach the asymptotic instability in a reasonable time.
However,
it must not be so high that it induces oscillations in the lock.
With a conventional two-stage integrating servo,
the ratio of the two gains must be no more than a few percent
(we use $\driftgain = \gain/50$ in our simulations).
Linear predictors optimised as in Sec.~\ref{sec:ctrldesign} are somewhat more robust against oscillations,
and can be operated with higher gain $\driftgain = \weight_1/10$ for the drift-correction integrator.

When post-processing measurement results to generate a virtual ``paper'' clock signal,
the causality requirements which limit the gain of the servo during physical clock operation no longer apply.
Thus, while the long-term stability limits discussed in Sec.~\ref{sec:ltstab} hold equally for physical and paper clocks because they arise from limits on the noise of the atomic reference,
the short-term stability limits discussed in this section can be avoided entirely in paper clocks and frequency ratio measurements,
where the LO frequency fluctuations can always be corrected as well as they can be measured.

\CHANGED{More abstractly, this section can also be understood in terms of the difference between steering the clock's frequency and steering its accumulated phase (i.e. indicated time).
The asymptotic limit of Eqn.~\ref{eqn:modsql} corresponds to an unavoidable random walk of phase due to the undetectable and uncorrelated frequency measurement errors of the atomic reference.
To reach it,
one must first correct the clock's output for all the detected LO frequency errors,
which dominate the short-term instability in multi-atom clocks.
Within our model this is done by the servo, the only component of the system with memory,
and thus the only component capable of remembering and correcting past phase errors:
an Allan deviation averaging down faster than $1/\sqrt{\tau}$ indicates that the servo is steering phase rather than simply locking frequency.
This can happen only slowly, however,
as it must not interfere with the servo's primary task of keeping the LO frequency near atomic resonance so that the reference continues to yield informative measurement results.
In clocks where separate corrections are applied to the output signal and to the signal used for atomic interrogation, the latter can be kept on resonance while the former's phase is corrected as fast as possible (even pre-emptively in the case of a paper clock),
thus minimising short-term fluctuations in the timing error.}

\section{Outlook}
\label{sec:outlook}

In this work we have studied the effects of LO noise on frequency standards
that monitor the LO frequency using a single ensemble of atoms
periodically interrogated according to a fixed protocol
and that correct the measured frequency fluctuations using a linear prediction formula.
Most current optical atomic clocks fit this description and can,
without hardware modifications,
use the framework presented here to identify and approach the stability limit imposed by their LO performance.
The interrogation times we recommend are specific to dead-time-free Ramsey interrogation,
but qualitatively similar results for other (Rabi, hyper-Ramsey) protocols can be found by the same arguments,
since our treatment of the servo is protocol-independent
and since all interrogation protocols face the same trade-off between measurement resolution
and unambiguous measurement domain.
Within this framework,
the most promising approaches to improving long-term clock instability
(besides improving LO performance)
seem to be those that improve the dynamic range of atomic measurements (such as spin squeezing),
whereas methods which attempt to make faster measurements with poor dynamic range (such as spectroscopy with maximally-correlated states)
have been shown to offer modest or no benefits for realistic LO noise spectra.

There are, however, many architectures for frequency standards that do not fit the framework presented here,
and it would be interesting to consider which of them can overcome the limits we have identified.
The simplest extension to implement would be the use of non-linear prediction algorithms,
which might improve the robustness of the servo,
allowing longer probe times and better stability at small atom number.
We expect that the performance of such algorithms would still be subject to the measurement-noise-independent limit of Eqn.~\ref{eqn:highNv},
so that they are unlikely to offer more than a modest constant-factor stability improvement in the large-$N$ limit.

Proposed multi-ensemble or cascaded clocks~\cite{Rosenband2013,Kessler2014}
circumvent the limits we have discussed here
by monitoring the LO noise with several different atomic references with progressively finer resolution.
References with a broad domain of useful frequencies provide coarse-resolution results
sufficient to narrow the prior $\priorvar$ for other, finer-resolution references.
The analysis we have presented here applies directly to the first (coarsest) reference in the cascade,
and the resulting stabilised signal can then be treated as an effective LO used by the next reference in the ensemble,
thus proceeding step-by-step down the cascade.
However, even if our analysis is locally valid for every reference treated individually,
the overall behaviour of such a multi-ensemble system may be qualitatively different than would be naively expected from the single-ensemble analysis \cite{Rosenband2013,Kessler2014,Chabuda2016}.

Finally, it would be interesting to make an analogous study for continuously-interrogated atomic references~\cite{Martin2011,Westergaard2015,Christensen2015,Chen2009,Meiser2009,Bohnet2012,Shiga2012,Norcia2016}.
Such systems,
whether based on continuous spectroscopic observation of an atomic sample
or on direct lasing on the clock transition (``active optical clock'')~\cite{Chen2009,Meiser2009}
face a conceptually similar trade-off between suppressing the noise of the atomic signal
(driving the system weakly to minimise the disturbance to the atoms)
and suppressing classical fluctuations in the probe laser, cavity mirrors, etc.
(driving the system strongly to gain information quickly and maximise the useful feedback bandwidth).
Thus,
the stability of these superficially different systems may depend on the noise of their classical components
and on the size of the atomic sample
in ways qualitatively similar to those we have examined here.

We thank S.~King and R.~Demkowicz-Dobrzański for stimulating discussions.
I.D.L. acknowledges a fellowship from the Alexander von Humboldt foundation.
The work presented here was partly supported through the project EMPIR 15SIB03 OC18.
This project has received funding from the EMPIR programme co-financed by the Participating States
and from the European Union's Horizon 2020 research and innovation programme.
We acknowledge support from the DFG through CRC 1128 (geo-Q), project A03 and CRC 1227 (DQ-mat), project B03.

\appendix

\section{Fluctuation correlation matrices \texorpdfstring{$\corr$}{C} for common noise processes}
\label{sec:explicitC}

In order to reconstruct LO noise properties from the experimentally observed correlation matrix $\corr$,
it is helpful to have explicit expressions for the correlations induced by common power-law noise processes,
which we summarise here.

Consider a continuous noisy process $\estfreq\of t$
with one-sided power spectral density $\spectrum\of f$,
whose autocorrelation reads
\begin{equation}
  \avg{\estfreq\of{t} \estfreq\of{t + \tau}} = \int_0^\infty\spectrum\of{f} \cos\of{2\pi f \tau} \dd f.
  \label{eqn:autocorr}
\end{equation}
If we associate the discrete estimates $\estfreq_j$ with time averages over a clock cycle of duration $\Tcyc$
such as would be measured by a perfect classical frequency counter,
\begin{equation}
  \estfreq_j = \frac{1}{\Tcyc}\int_{-j \Tcyc}^{(1 - j)\Tcyc} \estfreq\of{t} \dd t,
  \label{eqn:cycleavg}
\end{equation}
then the definition of Eqn.~\ref{eqn:defineC} reduces to
\begin{equation}
  \begin{split}
    \corr_{jk} = \frac{4}{\pi\Tcyc} \int_0^\infty \frac{\spectrum\of{u / \pi\Tcyc}}{u^2}
    & \sin^2\of{u} \sin\of{u j} \sin\of{u k} \\
    & \times \cos\of{u(j - k)} \dd u
  \end{split}
  \label{eqn:spectralC}
\end{equation}
where $u$ is a dimensionless dummy integration variable.

For a white-noise process [$\spectrum\of{f} \propto f^0$] of Allan deviation $\wndev$,
we find
\begin{align}
  \vec\corr =& \wndev^2\of\Tcyc \vec\wncorr \label{eqn:whiteCstructure} \\
  (\wncorr)_{jk} =& 1 + \delta_{jk} \label{eqn:whiteCgen} \\
  \vec\wncorr =& 
  \begin{pmatrix}
    2 & 1 & 1 & 1 & \cdots \\
    1 & 2 & 1 & 1 & \cdots \\
    1 & 1 & 2 & 1 & \cdots \\
    1 & 1 & 1 & 2 & \cdots \\
    \vdots & \vdots & \vdots & \vdots & \ddots
  \end{pmatrix}
  \label{eqn:whiteC}
\end{align}
as expected.
Detection noise of the atomic reference,
though not associated with the continuous frequency fluctuations of the LO,
yields the same white-noise correlation matrix.

For a flicker-noise process [$\spectrum\of{f} \propto f^{-1}$] of Allan deviation $\ffdev$,
\begin{align}
  \vec\corr =& \ffdev^2 \vec\ffcorr \label{eqn:flickerCstructure} \\
  (\ffcorr)_{jk} =& \difflog\of{\abs{j-k}} - \difflog\of{j} - \difflog\of{k}  \label{eqn:flickerCgen}\\
  \vec\ffcorr\approx& 
  { \renewcommand*{\arraystretch}{1.5}
    \begin{pmatrix}
      2 & 1.57 & 1.30 & 1.21 & \cdots \\
      1.57 & 3.13 & 2.43 & 2.08 & \cdots \\
      1.30 & 2.43 & 3.74 & 2.95 & \cdots \\
      1.21 & 2.08 & 2.95 & 4.16 & \cdots \\
      \vdots & \vdots & \vdots & \vdots & \ddots
    \end{pmatrix}
  }
  \label{eqn:flickerC}
\end{align}
where the auxiliary function $\difflog$ is defined as
\begin{equation}
  \difflog\of{n} = \frac{2\slog\of{n} - \slog\of{n-1} - \slog\of{n+1}}{4}
  \label{eqn:difflogdef}
\end{equation}
with
\begin{equation}
  \slog\of{n} =
  \begin{cases}
    n^2 \log_2 n & n>1 \\
    0 & \text{otherwise}
  \end{cases}
  \label{eqn:slogdef}
\end{equation}

For random-walk process [$\spectrum\of{f} \propto f^{-2}$] of Allan deviation $\rwdev$,
\begin{align}
  \vec\corr =& \rwdev^2\of\Tcyc \vec\rwcorr \label{eqn:walkCstructure} \\
  (\rwcorr)_{jk} =&  3\min\of{j, k} - \frac{1 + \delta_{jk}}{2}  \label{eqn:walkCgen} \\
  \vec\rwcorr=&
  { \renewcommand*{\arraystretch}{1.5}
    \begin{pmatrix}
      2 & \frac{5}{2} & \frac{5}{2} & \frac{5}{2} & \cdots \\
      \frac{5}{2} & 5 & \frac{11}{2} & \frac{11}{2} & \cdots \\
      \frac{5}{2} & \frac{11}{2} & 8 & \frac{17}{2} & \cdots \\
      \frac{5}{2} & \frac{11}{2} & \frac{17}{2} & 11 & \cdots \\
      \vdots & \vdots & \vdots & \vdots & \ddots
    \end{pmatrix}
  }
  \label{eqn:walkC}
\end{align}

Note that $\corr_{11}$ is, by definition, twice the single-cycle Allan variance for the noise process under consideration.
Also note that,
while Eqn.~\ref{eqn:spectralC} is valid only for perfect frequency-counting or dead-time free Ramsey interrogation,
Eqns.~\ref{eqn:whiteCstructure} through \ref{eqn:walkC},
expressed in terms of observed Allan variances,
are valid for arbitrary noisy time series of the specified power-law noise type,
including frequency estimates made with arbitrary measurement protocols that may include dead time.

\section{Optimal integrator gain with known noise}
\label{sec:optgain}
In simulations or in experiments where the LO is well-characterised,
it can be helpful to have an explicit formula for the optimal integrator gain in terms of known LO noise parameters.
To derive such a formula, we start by noting that for a pure white-noise process,
the variance of prediction errors for a simple integrator with gain $\gain$ is
\begin{equation}
  \wndev^2\of{\Tcyc} \frac{2}{2 - \gain},
\end{equation}
as can be shown by direct application of Eqn.~\ref{eqn:predvar} to the correlation matrix of Eqn.~\ref{eqn:whiteC}.
Similarly, for a pure random-walk process (and a hypothetical perfect classical reference),
the variance of prediction errors would be
\begin{equation}
  \rwdev^2\of{\Tcyc} \frac{3 - \gain}{\gain(2 - \gain)}.
\end{equation}
We have not found a similarly simple formula for the case of flicker noise,
but numerical studies shows that the phenomenological equation
\begin{equation}
  \ffdev^2\of{\Tcyc} \frac{1.6+0.4\gain-\ln4\ln\gain}{2 - \gain}
\end{equation}
is accurate to within \SI{2}{\percent} for gains in the range of \num{e-3} to 1.

As the integrator is a linear controller,
the mean-squared prediction error for a general noise process combining 
white, flicker and random-walk contributions is simply the sum of the three preceding expressions.
Differentiating this sum with respect to $\gain$
and imposing the condition $\gain >0$ to exclude servos that do nothing at all,
we find that the optimum gain
(the one which minimises the variance of prediction errors)
must satisfy
\begin{equation}
  \begin{split}
    &- 2\gain^2\ffvarratio\ln2\ln\gain \\
    &\quad+ \left[2 - \rwvarratio + (2.4 + \ln4)\ffvarratio\right]\gain^2
    + (6\rwvarratio - \ffvarratio\ln16)\gain
    - 6\rwvarratio
    = 0
  \end{split}
  \label{eqn:optgconstraint}
\end{equation}
where $\ffvarratio = \ffdev^2\of{\Tcyc} / \wndev^2\of\Tcyc$
and $\rwvarratio = \rwdev^2\of{\Tcyc} / \wndev^2\of\Tcyc$
are the flicker-floor and random-walk Allan variances at one cycle,
normalised to the total variance of all white noise contributions (including reference noise).
The first term on the left-hand side is always small and can be safely neglected,
leaving us with a quadratic equation in $\gain$.
Solving this equation yields a prescription for the gain
\begin{equation}
  \gain \approx \frac{\ffvarratio\ln4 - 3\rwvarratio + \sqrt{(\ffvarratio\ln4 - 3\rwvarratio)^2 + 6\rwvarratio a}}{a}
  \label{eqn:optgain}
\end{equation}
where we have introduced the usual auxiliary quantity
$a = 2 + (2.4 + \ln 4)\ffvarratio - \rwvarratio$.
Although approximate, Eqn.~\ref{eqn:optgain} yields integrating controllers
whose prediction error variance comes within \SI{1}{\percent} of that of the best numerically optimised integrators.

Also note that, in the absence of flicker- or random-walk noise,
the theoretically optimal integrator has a vanishingly small gain in order to average the white noise down as far as possible.
In practice, one should enforce a minimum acceptable gain,
in order to keep the servo time constant from growing unreasonable.
All our simulations enforce $\gain>0.04$,
i.e. a servo attack time no longer than 25 clock cycles,
though this bound is usually only relevant for unrealistic LO noise models or when the interrogation time of the atomic reference is too short.

\section{Posterior variance}
\label{sec:postvar}

Equation~\ref{eqn:linearvarred} gives the variance $\postvar$ of errors in the posterior estimate of the LO frequency
as a function of the variance $\priorvar$ of the servo prediction errors in the prediction
and of statistical properties of the measurement signal $\excfrac$.
Here we review the derivation of this posterior variance.

We wish to construct an estimate $\psi$ of the unknown phase $\phi$ corresponding to the error of the servo's prediction.
We take our estimate to be a linear combination of two (not necessarily independent) pieces of information: the expectation value $\avg\phi$,
which captures information available before the measurement result is revealed,
and the measurement result itself.
For later convenience, we express the measurement result as a deviation from the expectation value $\avg{\excfrac}$ of the measurement signal $\excfrac$. Thus:
\begin{equation}
  \psi = \alpha \avg{\phi} + \beta (\excfrac - \avg{\excfrac}).
\end{equation}
As in Sec.~\ref{sec:ltstab}, we define $\phi$ such that $\avg\phi = 0$.
The mean squared error $\postvar$ which we wish to minimise is then
\begin{align}
  \postvar =&\avg{(\psi-\phi)^2} = \avg{[\beta (\excfrac - \avg{\excfrac}) - \phi]^2} \\
  =& \beta^2 \avg{(\excfrac - \avg{\excfrac})^2} - 2\beta\avg{(\excfrac - \avg{\excfrac})\phi} + \avg{\phi^2}\\
  =& \beta^2 \var\of\excfrac - 2\beta\cov\of{\phi,\excfrac} + \priorvar  \label{eqn:minvar}
\end{align}
The minimum is attained for
\begin{equation}
  \beta = \frac{\cov\of{\phi, \excfrac}}{\var\of\excfrac}.
\end{equation}
Note that the weight $\beta$ given to the latest measurement result in estimating the frequency of the LO decreases as the measurement becomes noisier (i.e. as $\var\of\excfrac$ grows) or as the measurement becomes less strongly correlated with the underlying phase $\phi$ that we wish to estimate (i.e. as $\cov\of{\phi, \excfrac}$ shrinks).
The minimum posterior variance given in Eqn.~\ref{eqn:linearvarred} is obtained directly upon substitution of the optimised weight $\beta$ into Eqn.~\ref{eqn:minvar}.

For simplicity, we restrict ourselves to a \emph{linear} combination of the measurement signal $\excfrac$ with the prior estimate $\avg\phi$,
as this allows the results to be expressed entirely in terms of experimentally accessible (co)variances of noise distributions.
The variance of the errors in non-linear estimators depends on higher-order moments of the noise distributions which are difficult to characterise experimentally.
As argued at the end of Sec.~\ref{sec:ltstab},
non-linear estimators are empirically unnecessary, at least for simple Ramsey-like protocols.

\bibliography{ions}

\end{document}

%% file: macros.tex
\usepackage[T1]{fontenc}
\usepackage[utf8]{inputenc}
\usepackage[british]{babel} 
\usepackage{mathtools}
\usepackage[version=3]{mhchem}

\usepackage[full]{textcomp}
\usepackage{fourier}

\usepackage{gensymb}

\usepackage{bm}
\usepackage{siunitx}
\DeclareSIUnit\dBm{dBm}

\newcommand*\dd{\mathop{}\!\mathrm{d}} 

\newcommand*\ee{\mathrm{e}} 

\renewcommand*\vec[1]{\bm{#1}}

\newcommand*\tpose[1]{{#1}^\intercal}

\DeclareMathOperator\cov{cov}

\DeclareMathOperator\var{var}

\DeclarePairedDelimiter\abs{\lvert}{\rvert}
\DeclarePairedDelimiter\avg{\langle}{\rangle}

\DeclarePairedDelimiter\of{\lparen}{\rparen}
\DeclarePairedDelimiterX\ofgiven[2]{\lparen}{\rparen}%
{#1\,\delimsize\vert\,#2}

\DeclarePairedDelimiterX\comm[2]{\lbrack}{\rbrack}{#1,#2}

\DeclarePairedDelimiter\ket{\lvert}{\rangle}
\DeclarePairedDelimiterX\innerp[2]{\langle}{\rangle}{#1\,\delimsize\vert\,#2}
\DeclarePairedDelimiterX\matrixelt[3]{\langle}{\rangle}%
{#1\,\delimsize\vert\,#2\,\delimsize\vert\,#3}

\newcommand*\Sres{\ce{^{87}Sr}}
\newcommand*\Ybion{\ce{^{171}Yb+}}

\newcommand*\wnom{\omega}
\newcommand*\wlo{\omega_\mathrm{L}}

\newcommand*\lofreq{x}
\newcommand*\predfreq{h}
\newcommand*\errsig{e}
\newcommand*\estfreq{y}

\newcommand*\Tprobe{T}
\newcommand*\Tcyc{T_\mathrm{c}}
\newcommand*\Tnought{Z}
\newcommand*\TnoughtGHZ{Z_N}
\newcommand*\TnoughtGHZcyc{Z_{N\mathrm{c}}}
\newcommand*\Tcycnought{Z_{\mathrm{c}}}
\newcommand*\qpnnoise{\Delta\phi_\text{QPN}}

\newcommand*\Pexc{P_\mathrm{e}}
\newcommand*\prior{P}
\newcommand*\excfrac{F}

\newcommand*\priorvar{v}
\newcommand*\postvar{v^\prime}
\newcommand*\msmtvar{v_\mathrm{m}}

\newcommand*\loadev{\sigma_\mathrm{L}}
\newcommand*\clkadev{\sigma_\mathrm{C}}
\newcommand*\wndev{\sigma_\mathrm{w}}
\newcommand*\ffdev{\sigma_\mathrm{f}}
\newcommand*\rwdev{\sigma_\mathrm{r}}

\newcommand*\varprefactor\xi
\newcommand*\noisetype{\mu}
\newcommand*\ffvarratio{\beta}
\newcommand*\rwvarratio{\rho}
\newcommand*\stabconst{s}

\newcommand*\weight{w}
\newcommand*\Csize{n}
\newcommand*\gain{g}
\newcommand*\mainattack{\tau_1}
\newcommand*\driftgain{g_2}
\newcommand*\driftattack{\tau_2}

\newcommand*\spectrum{S}
\newcommand*\corr{C}
\newcommand*\wncorr{\corr_\mathrm{w}}
\newcommand*\ffcorr{\corr_\mathrm{f}}
\newcommand*\rwcorr{\corr_\mathrm{r}}
\newcommand*\difflog{\mathcal{D}}
\newcommand*\slog{\mathcal{L}}